\documentclass{article}

\usepackage[nonatbib, preprint]{neurips_2019}

\usepackage[utf8]{inputenc} 
\usepackage[T1]{fontenc}    
\usepackage{hyperref}       
\usepackage{url}            
\usepackage{booktabs}       
\usepackage{amsfonts}       
\usepackage{nicefrac}       
\usepackage{microtype}      

\usepackage{times}
\usepackage{epsfig}
\usepackage{graphicx}
\usepackage{amsmath}
\usepackage{amssymb}
\usepackage{paralist}
\usepackage{dsfont}
\usepackage{multirow}
\usepackage{wrapfig}
\usepackage{xcolor}
\usepackage{subfigure}
\usepackage{bm}
\usepackage{upgreek}
\input{taesup.sty}

\title{A Model-based Deep Learning Reconstruction for X-ray CT }

\author{%
  Kaichao Liang, Li Zhang, Yirong Yang, Hongkai Yang, Yuxiang Xing $^{*}$\\
  Key Laboratory of Particle \& Radiation Imaging, Tsinghua University\\
  Beijing, China 100084 \\
  $^{*}$Corresponding author: Yuxiang Xing  Email: xingyx@mail.tsinghua.edu.cn\\
}

\begin{document}

\maketitle

\begin{abstract}

Low dose CT is of great interest in these days. Dose reduction raises noise level in projections and decrease image quality in reconstructions. Model based image reconstruction can combine statistical noise model together with prior knowledge into an Bayesian optimization problem so that significantly reduce noise and artefacts. In this work, we propose a model-base deep learning for CT reconstruction so that a reconstruction network can be trained with no ground-truth images needed. Instead of minimizing cost function for each image, the network learns to minimize an ensemble cost function for the whole training set. No iteration will be needed for real data reconstruction using such a trained network. We experimented with a penalized weighted least-squares (PWLS) cost function for low dose CT reconstruction and tested on data from a practical dental CT.  Very encouraging results with great noise reductions are obtained.

\end{abstract}

\section{Introduction}

As X-ray CT is a widely used medical imaging modality for  clinical diagnosis tool, its radiation dose is of big concerned. High level of dose received by patients increase the risk of cancer \cite{Hall2008}. In the past decades, lowering X-ray dose by adjusting tube-voltage or tube-current-time have been studied extensively \cite{Naidich1990, Yoshinori2006, Yoshinori2005}. However, further reducing incident photons can cause high level statistical noise, and electronic noise becomes non-neglectable as well. If directly reconstructed with an FBP type algorithm, the noise in projection data will be magnified by the high-pass filtration (RL filter or SL filter) and degrade image quality.\\

Previously, many researches have been conducted on low dose CT (LDCT) image optimization\cite{Hall2008}. These algorithms generally can be divided into three categories.  The first category is filtration method. Manually designed filters or filtration algorithms are adopted to suspense noise. Pre-processing methods focus on projection domain denoising \cite{Jing2005Sinogram}, and post-processing methods focus on image domain denoising after analytical reconstruction \cite{Borsdorf2010Local}. Zhang et al. proposed a LDCT local texture restoration algorithm with priors extracted from NDCT image dataset \cite{Zhang2017Low}, which integrated NDCT reference dataset into filtration methods. However, low-pass filters restrain noise but generally cause detail information loss in projection data which further cause image blurring. The second category is iterative reconstruction methods \cite{Hao2015Adaptive, Zhang2014Statistical, Qiong2012Low}. Model-based iterative reconstruction (MBIR) is based on Bayesian theory and models noise model, physical process and system configuration into reconstruction. For example, combining statistical iterative reconstruction (SIR) with none-local-mean (NLM) constraint for low-dose CT reconstruction \cite{Hao2015Adaptive, Zhang2014Statistical}  to suppress noise.  Instead of using NLM, Xu adopted dictionary learning which learns prior constraint based on normal-dose CT dataset \cite{Qiong2012Low}. The images reconstructed from low-dose projections are expected to be of similar distribution as normal dose CT reconstructions. However, iterative process is very time consuming which is obstructed its usage in real-time reconstruction. Besides, hyper-parameters in cost function and number of iteration need to be tuned according to different data.\\

Recently, deep learning methods have been widely conducted in LDCT problems \cite{Wang2017Low, Chen2017Low, Zhang2018CT, Yang2017Low}. Hu and his colleagues used a three-layer CNN structure for LDCT image optimization and achieved visible image quality improvements \cite{Wang2017Low}. Later, Chen proposed a more complex encoder-decoder structure CNN for LDCT and achieved better results \cite{Chen2017Low}.  Convolutional neural networks (CNN) learns the relationship between LDCT images and NDCT images from training set.  Because of the large cost of real data, LDCT and corresponding NDCT images are often from simulations. In these works, normal dose CT (NDCT) images are required to form labels, which is easy to access in simulation study. However, in practical complex situation, features of real data can be very different from simulation, which could heavily influences the generalization of networks. No ground-truth images available for labels in real situation become an obstacle for practical problems.\\

In this work, we propose a model based deep learning (MBDL) method for CT reconstruction that requires no ground-truth information. The network can be directly trained with LDCT projection data only, encouraging reconstruction results from practical low dose dental CT data are obtained.

\section{Methods}

\subsection{Cost function to model X-ray CT imaging}

In regression problem, labels are usually used to form L1/L2-norm as the cost function of networks to minimize. In many situations, ground-truth information might be available. The basic idea of our method is to incorporate a system model based cost function for deep-learning network. Without losing the generality, we took Bayesian estimation principle to construct our cost function for CT reconstruction similar to MBIR methods. In general, Bayesian cost function contains two terms: data fidelity term which models the noise of projection data, and prior term which constraint the reconstructed image to follow prior information about general image characteristics.\\
\begin{equation}
\Phi\left(\bm{\upmu}\right) = \phi_{\rm f} \left(\bm{\upmu},\textbf{p}, \textbf{H}\right)+\lambda\phi_{\rm p}\left(\bm{\upmu}\right)\label{eq1}
\end{equation}
Here, $\Phi$ denotes a general Bayesian cost function with $\phi_{\rm f}$ the data fidelity term and $\phi_{\rm p}$ the priors. Here, $\textbf{p}$ is the a CT projection data that is very noisy from a low dose scan, and $\textbf{H}$  the system matrix of CT, $\bm{\upmu}$ denotes the estimation of linear attenuation map, i.e. be the output of network. The hyper-paramete $\lambda$ in Eq.\ref{eq1}  is  commonly used to adjust the strength of two terms. Hence, with known imaging system matrix and projection data, the cost function avoids using any ground-truth label and model imaging physics instead.\\

In X-ray CT imaging, the data are generally modelled to be bin-to-bin independent Poisson or Compound Poisson distributed\cite{Elbakri2003Efficient}. After negative log operation of Poisson distributed data, a sinogram projection is approximately Gaussian with its variance dependent on mean \cite{Wang2008An}. Thus, we have the likelihood to be:
\begin{equation}
\textbf{P}\left(\textbf{p};\bm{\upmu}\right) = \exp\left(-(\textbf{H} \bm{\upmu}-\textbf{p}\right)^T\bm{\Sigma}^{-1} \left(\textbf{H} \bm{\upmu}-\textbf{p})\right)\label{eq0}
\end{equation}
Here, $\bm{\Sigma}^{-1}$ is a diagonal matrix and its diagonal elements denote the variance of projections. Various priors have been studied in X-ray CT reconstruction \cite{Zhang2014Statistical, Qiong2012Low, Ludwig2011Improved, Kyungsang2015Sparse, Zhang2016A}. In this work, we adopt a constraint derived from none local mean (NLM) [8]. 
The NLM term can be formulated as:
\begin{equation}
\phi_{\rm NLM}\left(\bm{\upmu}\right)=\frac{1}{2}\Sigma_{i}\Sigma_{i'\in \mathcal N\left(i\right)}W_{ii'}\left(\mu_{i}-\mu_{i'}\right)^2 \label{eq3}
\end{equation}
Here, $\mathcal N\left(i\right)$ denotes the neighboring pixels of $i$ within a searching window and $W_{ii'}$ denotes the weight of similarity computed from:
\begin{equation}
W_{ii'}=\frac{{\rm exp}\left(-||\Pi\left(\mu_{i}\right)-\Pi\left(\mu_{i'}\right)||_{2,a}^{2}/h\right)}{\Sigma_{j \in \mathcal N \left(i\right)}{\rm exp}\left(-||\Pi\left(\mu_{i}\right)-\Pi\left(\mu_{j}\right)||_{2,a}^{2}/h\right) }\label{eq4}
\end{equation}
With $\Pi\left(\cdot\right)$ being a patch extractor, $||\cdot||_{2,a}$ the L2-norm with a Gaussian kernel $a$ to weight the distance in a patch and a normalization parameter. If neighboring patches of two pixels are similar, the two pixels are likely to be the same kind of materials and gets a high weight. In this way, NLM can preserve structures well while restrain noise.
Combining Eq. \ref{eq0} and \ref{eq3}, we obtain a Bayesian cost function which falls into the category of penalized weighted least square (PWLS). It is used to guide our deep learning network:
\begin{equation}
\Phi\left(\bm{\upmu}\right) = \left(\textbf{p}-\textbf{H}\bm{\upmu}\right)^T\bm{\Sigma}^{-1}\left(\textbf{p}-\textbf{H}\bm{\upmu}\right)+\lambda\phi_{\rm NLM}\left(\bm{\upmu}\right) \label{eq2}
\end{equation}\\

\subsection{Network architecture and training}

\begin{figure}[H]
\centerline{\includegraphics[width = 0.92\linewidth]{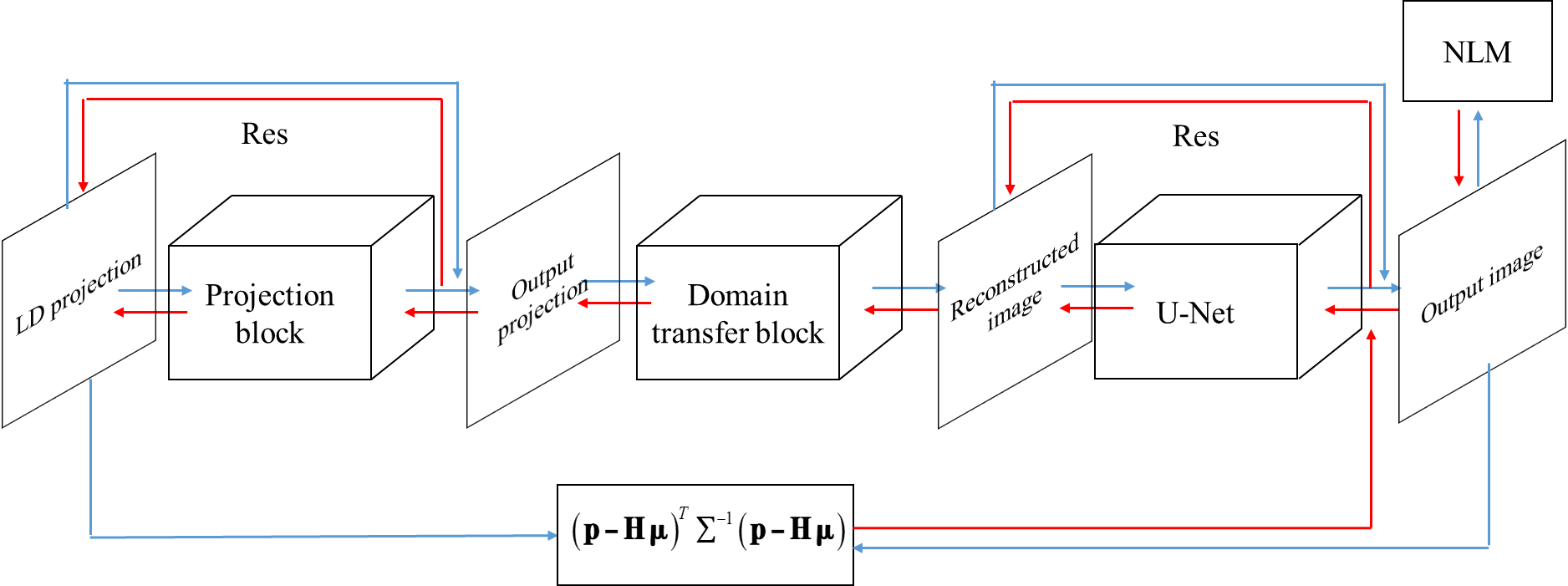}}
\caption{Architecture of model-based deep learning network for CT reconstruction.}
\label{fig1}
\end{figure}

To enforce the cost function as in Eq.\ref{eq2}, we constructed a reconstruction network as displayed in Fig. \ref{fig1}. It takes noisy projection data as inputs and output reconstructed images at the end. The network is consisted of three blocks to do dual-domain learning similar to our previous work \cite{Liang2018Comparison, Yang2019Slice}.\\

First, projection block reduces the noise in projection domain. Projection block contains five convolutional layers. Each convolutional layer is followed with leaky-Relu activation function. Next, domain transfer block is an analytical inverse operator. We adopt the matrix format of FBP reconstruction as Eq.\ref{eq5}
\begin{equation}
\bm{\upmu}=\textbf{H}_W^T \textbf{F} \left( \textbf{W $\cdot$ p} \right)
\label{eq5}
\end{equation}
Here, \textbf{W} denotes projection weighting matrix, it weights the projection. \textbf{F} is a matrix form of RL filtration, it filters the projection data along the detector axis. $\textbf{H}_W^T$  denotes the operator of weighted back-projection. All the matrix in the reconstruction block is pre-calculated and not trained in this study, but gradients can path through the reconstruction block from image domain back to projection domain, which is critical in our network. Following reconstruction block, an image domain U-Net \cite{Ronneberger2015U} refines the reconstruction results and output the final estimation. The U-Net structure achieves big receptive-field with pooling layers, while it recovers local spatial resolution by combining up-sampled features with un-pooled features. Both image domain network and projection domain network adopt residual structure to ease the network training process.\\

In our training, the network parameters are updated to minimize the Bayesian cost function over the whole training set.
\begin{equation}
v_1,...,v_M=\mathop{\rm argmin}\limits_{v_1,...,v_M}\Sigma_{k\in1:K}\Phi\left(\bm{\upmu}^k\right)
\label{eq6}
\end{equation}
Here, $k$ indexes the attenuation map in the training dataset of size $K$, and $v_1,...,v_m,...,v_M$ denote the parameters in the network.\\

For each training step, gradients of cost function with respect to $k^{th}$ output image can be calculated from Eq. \ref{eq2} as:
\begin{equation}
\frac{\partial \Phi\left(\bm{\upmu}^k\right)}{\partial \mu_{i}^{k}}=\left[\textbf{H}^{T}\left(\textbf{H}\bm{\upmu}^k-\textbf{p}\right)\right]_i\bm{\Sigma}_{i}^{-1}+\lambda\Sigma_{i'\in\mathcal N\left(i\right)}W_{ii'}\left(\mu_{i}^{k}-\mu_{i'}^{k}\right)
\label{eq7}
\end{equation}
Here, subscript $i$ denotes the $i^{th}$ element of a vector.  Further, the gradients of the cost function with respect to network parameters can be calculated by chain rule:
\begin{equation}
\frac{\partial \Phi\left(\bm{\upmu}^k\right)}{\partial v_{m}^{n}}=\Sigma_{i}\frac{\partial \Phi\left(\bm{\upmu}^k\right)}{\partial \mu_{i}^{k}} \cdot \frac{\partial \mu_{i}^{k}}{\partial v_{m}^{n}}
\label{eq8}
\end{equation}
with $n$ indexing the $n^{th}$ training iteration. \\

Parameters are updated with standard gradient descent method to minimize the cost function:
\begin{equation}
v_{m}^{n+1}=v_{m}^{n}-\alpha\Sigma_{k\in1:K}\frac{\partial \Phi\left(\bm{\upmu}^k\right)}{\partial v_{m}^{n}}
\label{eq9}
\end{equation}
Here, $\alpha$ denotes the learning rate. Other optimization method can also be used to replace the gradient descending method. It is worth to emphasize that we train the network by directly updating parameters of network as above.\\

Different from supervised pixel-to-pixel loss, Bayesian cost function does not provide clear optimization direction in image domain for output images. To be more specific, the gradients of output image do not force output image with a clear goal in image domain in different training steps, which makes the network converges slowly in training. Thus, we applied some tricks to accelerate training process. In the early epochs, we train each data in training set for several times continuously, then change to another training data. Training each data continuously help network to catch more robust rule from Bayesian cost function. Moreover, in the early epochs, we only train the network with data fidelity term so that the main structure of reconstructed images can be learned efficiently. After the pre-training process, prior constrain is added.\\

\section{Experiments and results}

\subsection{Study with simulated projection data}

We first carried out our experiments on 3872 practical high quality dental CT images. We simulated the LDCT projections as network inputs. Among them, 3400 randomly chosen simulated projection data were used as training set while the left projections and the corresponding high-quality images formed test set. The high-quality dental CT images were not available to training process but only used as ground-truth for performance evaluation of the reconstruction results during inference, thus the results can be evaluated both in vision and quantitatively. Both ground-truth images and reconstruction images were $640^2$ with pixel size 0.25mm. The projections were of uniformly distributed 720 fan-beam views over 2$\pi$, and 640 detector bins with bin size 0.512mm. The projections were simulated with ray-tracing forward projection method and Poisson noise was added with 5000 incident photons.\\

\begin{figure}[p]
\centering
\includegraphics[width = 1\linewidth]{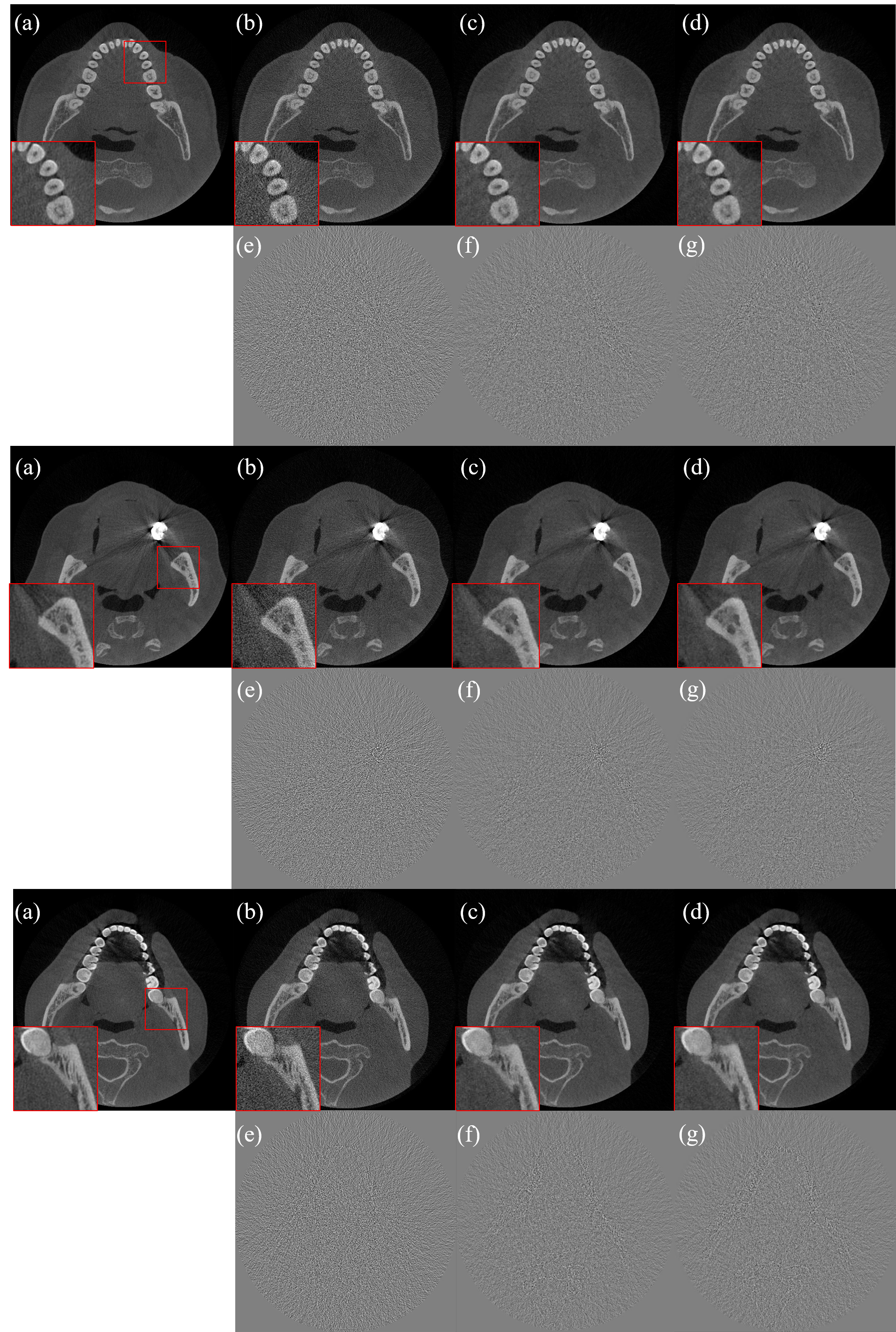}
\caption{Comparison of different reconstruction methods. (a): Ground-truth, (b):  LDCT FBP, (c): NLM-WLS iterative reconstruction, (d): LDCT MBDL reconstruction, (e): Residual map between (a) and (b), (f): Residual map between (a) and (c), (g):  Residual map between (a) and (d).}
\label{fig2}
\end{figure}

In Fig.\ref{fig2}, ground-truth, LDCT FBP reconstruction, NLM-WLS iterative reconstruction, and MBDL network reconstruction were displayed for visual comparison. Details denoted by red boxes were zoomed-in and displayed in the lower left corner.  As shown in Fig.\ref{fig2}, the noise was well suppressed in both iterative reconstruction and MBDL reconstruction, the details of teeth and bone were restored to a large extent as in zoomed-in.

For further quantitative analysis, we compared different LDCT reconstruction results with ground-truth images by calculating mean RRMSE and SSIM indexes based on 25 randomly chosen test images:
\begin{equation}
{\rm RRMSE}\left(\bm{\hat{\upmu}}, \bm{\upmu}\right)=\frac{||\bm{\hat{\upmu}}-\bm{\upmu}||_2}{||\bm{\upmu}||_2}
\label{eq10}
\end{equation}

\begin{equation}
{\rm SSIM}\left(\bm{\hat{\upmu}}, \bm{\upmu}\right)=\frac{\left(2\bm{\hat{\upmu}}\bm{\upmu}+C_1\right)\left(2\sigma_{\bm{\hat{\upmu}},\bm{\upmu}}^2+C_2\right)}{\left(\bm{\hat{\upmu}}^2+\bm{\upmu}^2+C_1\right)\left(\sigma_{\bm{\hat{\upmu}}}^2+\sigma_{\bm{\upmu}^2}^2+C_2\right)}
\label{eq11}
\end{equation}

Here, $\bm{\hat{\upmu}}$ denotes LDCT reconstruction results of different methods to be evaluated, and $\bm{\upmu}$ corresponding ground-truth. $C_1, C_2$ are constants. The mean and standard deviation of RRMSE and SSIM indexes between different FBP reconstruction methods and ground-truth were compared in Tab.\ref{tab1}. Both WLS-NLM iterative reconstruction and MBDL reconstruction significantly reduce the RRMSE and improve SSIM indexes from LDCT FBP reconstruction results, while the MBDL reconstructions achieved slightly better quantitative performance than WLS-NLM iterative reconstructions in practice.

\begin{table}[H]
\caption{ RRMSE and SSIM between gorund-truth and LDCT reconstruction}
\begin{center}
\begin{tabular}{|c|c|c|c|}
\hline
\textbf{Method} & LDCT FBP & LDCT WLS+NLM IR & LDCT MBDL\\
\hline
\textbf{RRMSE} & 0.2560$\pm$0.0094 & 0.1164$\pm$0.0022 & 0.1082$\pm$0.0039\\
\hline
\textbf{SSIM}  & 0.9394$\pm$0.0083 & 0.9868$\pm$0.0015 & 0.9885$\pm$ 0.0015\\
\hline
\end{tabular}
\end{center}
\label{tab1}
\end{table}

\subsection{Reconstruction of practical raw data with additional simulated noise}

\begin{figure}[H]
\centering
\includegraphics[width = 1\linewidth]{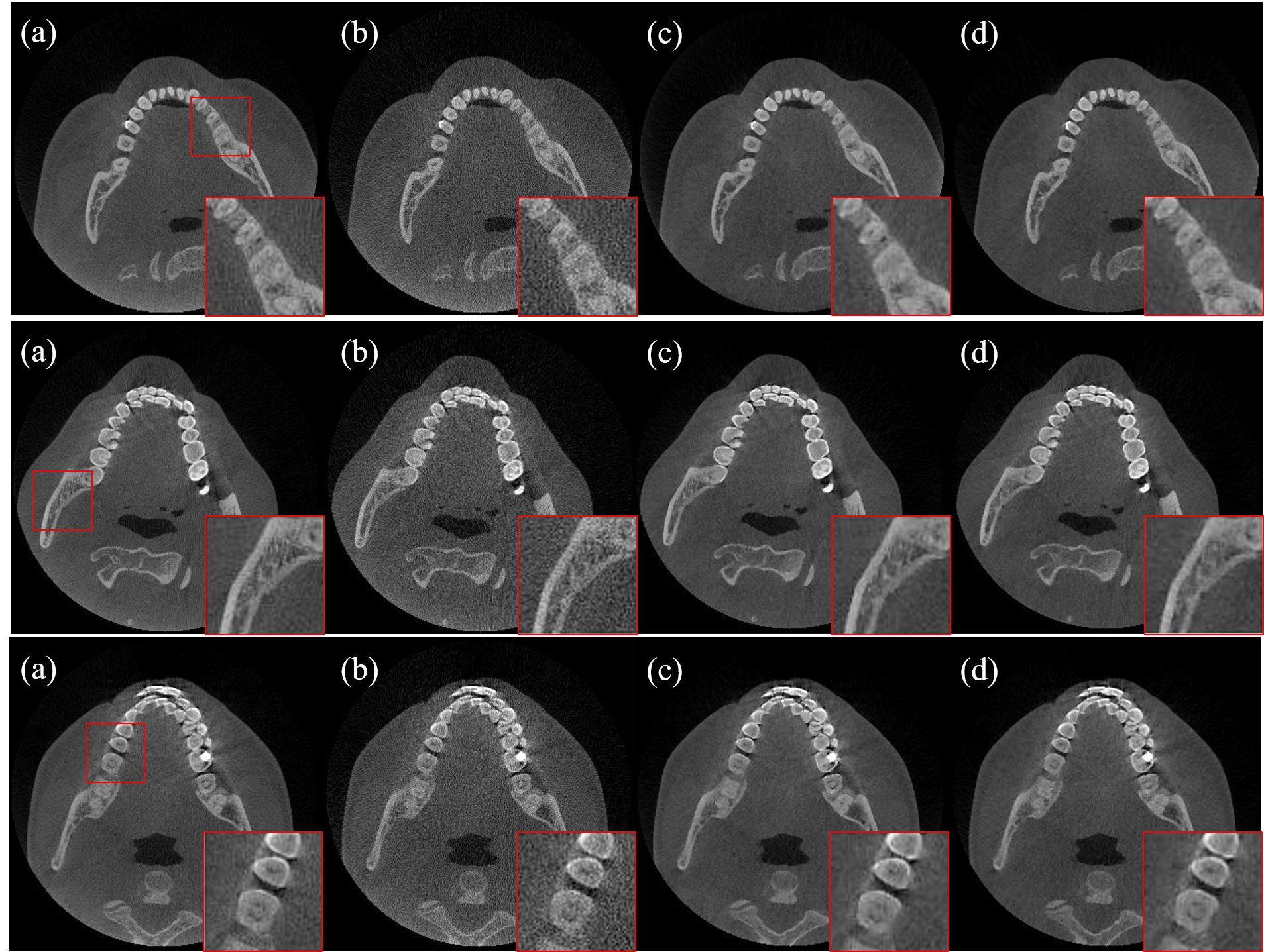}
\caption{Comparison of different reconstruction methods. (a): NDCT Hanning + FBP, (b):  LDCT Hanning + FBP, (c): NLM-WLS iterative reconstruction, (d): LDCT MBDL reconstruction.}
\label{fig3}
\end{figure}

We further carried out our experiments on practical dental CT raw data while manually increasing the noise level by additional Poisson noise, thus the situation is close to reality LDCT. In total, raw data from 3872 patients were collected. Among them, randomly chosen 3400 data were used as training set with the others forming test set for the proposed network reconstruction method. The projections were of uniformly distributed 600 fan-beam views over 2$\pi$, and 656 detector bins covering half of the ROI (reduced size detector) \cite{Li2007A}. The size of detector bins was 0.25mm. To simulate low dose situation, original normal-dose projection data (no image domain ground truth) were treated as noise-free reference data, Poisson noise was added with 5000 incident photons. Reconstruction images were $360^2$ with pixel size 0.5mm.\\

In Fig.\ref{fig3}, FBP reconstruction of NDCT with Hanning filter, LDCT FBP reconstruction with Hanning filter, NLM-WLS iterative reconstruction, and MBDL network reconstruction were displayed for visual comparison. We applied a 5-order Hanning window filtration in FBP reconstructions for noise reduction in this comparison. Details denoted by red arrow were zoomed-in and displayed in the lower right corner. Similar with full-simulation cases, both iterative reconstruction and MBDL network reconstruction suppressed the noise and restored rich structures of teeth. The two algorithms achieved similar visual results.

\section{Conclusion}
In this work, we proposed a model-based deep learning reconstruction for LDCT. The proposed method adopt a Bayesian cost function and construct a dual-domain network allowing easy error propagation for the cost function minimization. The MBDL network is trained to minimize the cost function for the whole dataset with no ground-truth information needed. After training, the network can complete reconstruction without time-consuming iterative steps and reduced noise in a level comparable to iterative methods. Fundamentally, the proposed method integrates the optimization principle of iterative reconstruction methods with deep learning. This method can be easily extended to other scan modality and geometry with constraints or various combination of regularizations for problems such as sparse-view and limited-angle CT reconstruction.

\newpage

\bibliographystyle{unsrt}
\bibliography{MBDL_LDCT.bib}

\end{document}